\documentclass{article}
\usepackage{spconf,amsmath,graphicx}
\usepackage{multirow}
\usepackage{amsfonts}
\usepackage{bbold}
\usepackage{array}
\usepackage{soul}
\usepackage{graphicx}
\usepackage{pgfplots}
\usepackage{url}
\usepackage{ tikz }
\usepackage{silence}
\usepackage{cite}
\usepackage{mathtools}
\usepackage[inline]{enumitem}
\usepackage{accents}

\ninept

\title{Stabilising and accelerating light gated recurrent units \\ for automatic speech recognition}
\name{Adel Moumen$^{1}$, Titouan Parcollet$^{1,2}$}
\address{$^{1}$Laboratoire Informatique d'Avignon, Avignon Université \\ $^{2}$Cambridge Machine Learning Systems Lab, University of Cambridge}
\begin{document}
%

\maketitle

\begin{abstract}
The light gated recurrent units (Li-GRU) is well-known for achieving impressive results in automatic speech recognition (ASR) tasks while being lighter and faster to train than a standard gated recurrent units (GRU). However, the unbounded nature of its rectified linear unit on the candidate recurrent gate induces an important gradient exploding phenomenon disrupting the training process and preventing it from being applied to famous datasets. In this paper, we theoretically and empirically derive the necessary conditions for its stability as well as engineering mechanisms to speed up by a factor of five its training time, hence introducing a novel version of this architecture named SLi-GRU. Then, we evaluate its performance both on a toy task illustrating its newly acquired capabilities and a set of three different ASR datasets demonstrating lower word error rates compared to more complex recurrent neural networks.


\end{abstract}
\section{Introduction}
\label{sec:intro}


Automatic Speech Recognition (ASR) is the task of transforming spoken language into text. Myriads of real-life devices and applications ranging from personal assistants to smart cars and automatic captioning rely on advanced ASR systems to provide the optimal experience to end users. ASR has greatly benefited from the rise of deep learning to improved robustness and performance. In practice, various deep learning techniques and models exhibiting different strengths and weaknesses have been developed over the last decade. Transformer neural networks, for instance, have reached unprecedented levels of word error rates for offline ASR \cite{karita2019comparative}. Sequence-to-sequence (seq2seq) modelling with recurrent neural networks (RNN), on the other hand, has steadily achieved top-notch performance both for online and offline ASR while remaining fairly simpler to approach and implement both for academia and for the industry. Nowadays, transformers and RNN still co-exist widely, and the final choice of the architecture boils down to the desired use-case and the available compute resources. A successful implementation of a recurrent CTC-Attention encoder-decoder was first proposed in \cite{watanabe2017hybrid}, and consisted in a recurrent encoder with an attentional recurrent decoder trained jointly with the CTC loss function \cite{graves2006connectionist, graves2014towards}. Such encoder-decoder architectures offer a rich literature as well as an ever-growing number of open-source implementations \cite{ravanelli2021speechbrain,watanabe2018espnet} which continue to power many real-life products relying on ASR.

Following the CTC-Attention paradigm, one must find the most appropriate encoder to turn the speech signal into a latent subspace that the decoder can further process. In practice, and based on the successful DeepSpeech architecture \cite{amodei2016deep}, many competitive encoders simply combine convolutional layers with a deep recurrent neural network. Hence, the choice of the recurrent unit is of crucial interest to achieve state-of-the-art word error rates. For instance, the light gated recurrent units (Li-GRU) \cite{ravanelli2018light} network has been designed to carefully address the task of ASR. A Li-GRU is a compact single-gate unit derived from the gated recurrent units (GRU) which reduce by $30\%$ the per-epoch training time over a standard GRU while also improving the ASR accuracy. Nevertheless, and despite a clear interest from the community, two major issues prevent a stronger adoption of the Li-GRU: (1) it highly suffers from exploding gradients as the gate is unbounded; and (2) no optimized implementation exists, hence leading to much larger training times than more complex alternatives such as LSTM neural networks. 


The problems of vanishing and exploding gradients in recurrent neural networks (RNNs) have been studied for decades and several methods have been proposed. Gradient clipping and weight decay, for instance, directly act on the values that are flowing throughout the network during the forward and backward propagations \cite{loshchilov2017decoupled,pascanu2013difficulty}. Regularization techniques, such as the soft orthogonal regularisation (SOR) \cite{vorontsov2017orthogonality} constraint, propose to directly tackle the problem of uncontrolled growth of the error accumulation during backpropagation through time (BPTT) \cite{werbos1990backpropagation} by introducing a term in the loss enforcing the recurrent weight matrices to remain orthogonal \cite{vorontsov2017orthogonality}. Others even proposed to develop a novel RNN architecture learning a unitary recurrent weight matrix in the complex domain \cite{arjovsky2016unitary}. An even more stable approach is to directly clip the eigenvalues of the recurrent weight matrices as soon as they reach a certain threshold \cite{NIPS2017_f2fc9902}. However, the two latest solutions introduce important complexity overheads at computation time and remain mostly intractable for ASR scenarios with medium and large datasets. Then, and despite promising research directions, the instability of the Li-GRU has never been formally or empirically investigated. It remains impossible to verify if existing methods preventing exploding gradients could benefit the Li-GRU. 

%
%



This article proposes to formalise and solve the issue that the original Li-GRU encounters. Hence, it introduces the Stabilised Li-GRU (SLi-GRU), offering theoretical and empirical guarantees on its stability compared to the original Li-GRU as well as a five times reduced training time following a well-designed CUDA implementation. In summary, the contributions are fourfold: 1. Theoretically ground the conditions to avoid exploding gradients with the Li-GRU; 2. Analyse the impact of existing methods both theoretically and empirically with the additive task on the Li-GRU, leading to the introduction of the SLi-GRU; 3. Empirically validate the scaling of the SLi-GRU to three well-known speech recognition tasks; and 4. Deliver to the community a CUDA optimized version of the Li-GRU and SLi-GRU within SpeechBrain\footnote{https://speechbrain.github.io/}.

%
%


The conducted experiments on LibriSpeech and CommonVoice demonstrate that the SLi-GRU is much more stable and faster than the original Li-GRU which can not be trained with such datasets. It also obtains better ASR performance than the state-of-the-art baseline offered within Speechbrain based on LSTM neural networks.

\section{On the Instability of the Li-GRU}
\label{sec:pge}

The initial definition of the Li-GRU \cite{ravanelli2018light} is unstable and, in practice, can not be applied to even medium-sized speech recognition datasets such as Librispeech or CommonVoice. First, we recall the basic equations of the Li-GRU (Section \ref{subsec:ligru}). Then, following a theoretical analysis of the gradient exploding phenomenon arising at training time, we derive the necessary bounds to ensure a smooth training (Section \ref{subsec:analytical}). Finally, we formalise various approaches to tackle the gradient exploding problem by investigating theoretically their impact on the later boundaries (Section \ref{subsec:methods}). 


\subsection{Light gated recurrent units}
\label{subsec:ligru}

The Li-GRU \cite{ravanelli2018light} has been designed to carefully address the task of speech recognition efficiently. By removing the reset gate, adding a ReLU on the candidate gate, and applying batch normalization ($\mathcal{BN}$) \cite{ioffe2015batch} on the feed-forward connections the authors were not only able to reduce by $30\%$ the per-epoch training time of a Li-GRU against a standard GRU \cite{cho2014properties}, but also to improve the performance across different tasks. In particular, the Li-GRU equations are:
\begin{equation}
    \label{eq:zt_ligru}
    z_{t} = \sigma(\mathcal{\mathcal{BN}}(W_{z}x_{t}) + U_{z}h_{t-1}),
\end{equation}
\begin{equation}
    \label{eq:relu_ligru}
    \tilde{h} = \text{ReLU}(\mathcal{\mathcal{BN}}(W_{h}x_{t}) + U_{h}h_{t-1}),
\end{equation}
\begin{equation}
    h_{t} = z_{t} \odot h_{t-1} + (1 - z_{t}) \odot \tilde{h}.
\end{equation}
with $z_{t}$ the update gate, $\tilde{h_{t}}$ the candidate gate, $h_{t}$ the hidden state, all taken at the time step $t$, and $h_{t-1}$ the hidden state from the previous time step. The logistic sigmoid function is denoted as 
$\sigma$ while the operator $\odot$ refers to the element-wise product. The Li-GRU is fed at each time step with a vector $x_{t}$. $ {W}_{z}$, $ {W}_{h}$  are the feed-forward, and $ {U}_{z}$, $ {U}_{h}$ the recurrent weights. In the following, we consider $ {W}_{*}$ and $ {U}_{*}$ as references to the weights independently of the gates.


\subsection{Gradient instabilities and temporal contributions}
\label{subsec:analytical}

Despite its simplicity, the Li-GRU exhibits a weakness in Eq. \ref{eq:relu_ligru} as the recurrent process is unbounded, and therefore subject to potential gradient instabilities. 
Here, and following carefully previous work from \cite{pascanu2013difficulty,NEURIPS2019_f8eb278a,vorontsov2017orthogonality, arjovsky2016unitary}, we propose to study from an analytical point of view the theoretical motivations of the Li-GRU instability.  

Let $E = \sum_{N}^{t=0} E_{t}$ be the total loss, where $N$ is the sequence length, and $E_{t}$ the loss at the time step $t$. We can highlight the gradient exploding problem in the Li-GRU with the BPTT by showing that each backpropagation step in the recurrent process is controlled by a value $\eta$ that may explode as soon as it deviates from $1$. Following the BPTT, we first obtain $\partial  E_{t}/\partial h_{t}$ and then backpropagate from $\partial  E_{t}/\partial h_{m}$ to $\partial  E_{t}/\partial h_{m-1}$ with ($m \leq t$):
\begin{equation}
    \label{eqn:error_in_time_1}
    \frac{\partial E_{t}}{\partial h_{m}} = \frac{\partial E_{t}}{\partial h_{t}}(\prod_{i=m}^{t} \frac{\partial h_{i}}{\partial h_{i-1}}),
\end{equation}
\begin{equation}
    \label{eqn:error_in_time_2}
    \frac{\partial E_{t}}{\partial h_{m}}  = \frac{\partial E_{t}}{\partial h_{t}}(\prod_{i=m}^{t}\frac{\partial ^+ h_{i}}{\partial h_{i-1}} + \frac{\partial h_{i}}{\partial z_{i}} \frac{\partial z_{i}}{\partial h_{i-1}} + \frac{\partial h_{i}}{\partial \tilde{h_{i}}} \frac{\partial \tilde{h_{i}}}{\partial h_{i-1}}),
\end{equation}
with $\partial ^+h_{i}/\partial h_{i-1}$ refering to the immediate partial derivative. Then, we derive the $\partial h_{i}/\partial h_{i-1}$ upper bound called $\eta$ that will describe the necessary conditions triggering the explosion phenomenon. $\eta$ is obtained by computing the norm of each partial derivatives of $\partial h_{i}/\partial h_{i-1}$:
\begin{equation}
    \label{eqn:eta}
    \eta = \frac{\gamma_{1}}{4} || {U}_{z}||_{2} + || {U}_{h}||_{2},
\end{equation}
with $\gamma_{1} = max_{1 \leq m \leq t, 1 \leq j \leq d} | [h_{m-1}]_{j}|$, and $d$ the hidden dimension size. Then, considering $||\partial E_{t}/\partial h_{m-1}||$ and $||\partial E_{t}/\partial h_{m}||$ at adjacent timesteps and leveraging the norm properties:
\begin{equation}
    || \frac{\partial E_{t}}{\partial h_{m-1}} || \leq \eta || \frac{\partial E_{t}}{\partial h_{m}} ||.
\end{equation}
Finally, by induction, we generalize over non adjacent time steps noted $p$ subject to $p < m$ (\textit{i.e} applying $\eta$, $m - p$ times):
\begin{equation}
    || \frac{\partial E_{t}}{\partial h_{p}} || \leq \eta^{m - p} ||\frac{\partial E_{t}}{\partial h_{m}}||.
\end{equation}
It results that the Li-GRU may explode as soon as $\eta > 1$ following the increase of the $m-p$ quantity, hence, leading to the exploding gradients phenomenon. Therefore, we wish to keep $\eta$ as close as possible to $1$.

\subsection{Stabilizing the Li-GRU}\label{subsec:methods}
This section presents various methods to tackle exploding gradients in the Li-GRU by analysing their impacts on $\eta$.\\

%
%


\noindent\textbf{Soft orthogonal regularization (SOR).} The SOR \cite{vorontsov2017orthogonality} helps any RNN to keep $ {U}_{*}$ close to orthogonal by adding a regularization term to the loss, hence limiting the impact of $ {U}_{*}$ on $\eta$. The spectral norm of a matrix $||{W}||_{2}$ is $1$, simplifying the value of $\eta$ from Eq. \ref{eqn:eta}:
\begin{equation}
    \eta = \frac{\gamma_{1}}{4} + 1.
\end{equation}
Despite being helpful to approach the problem of exploding gradients, such a regularization will constrain the weights to be in the Stiefel manifolds, and potentially degrade the global training loss.\\


\noindent\textbf{Weight decay and gradient clipping.} Weight decay (WD) \cite{loshchilov2017decoupled} adds a penalty on the l2-norm of $ {U}_{*}$ and $ {W}_{*}$ potentially reducing the impact of the recurrent weights on $\eta$. Nevertheless, this method is not always sufficient and often is coupled with gradient clipping (GC) \cite{pascanu2013difficulty, mikolov2012statistical} that rescales the gradient so that its norm remains lower than a threshold, hence postponing the growth of $ {U}_{*}$. In practice, this method is effective for ${U}_{h}$ that is never rescaled. \\

\noindent\textbf{Sine activation function.} In \cite{parascandolo2016taming}, the authors demonstrated that the sine activation function was an excellent bounded alternative for recurrent neural networks. In our case, it will be the non-linearity of the candidate gate and therefore modify the expression of $\eta$ as:
\begin{equation}
    \eta = \frac{\gamma_{1}}{4} || {U}_{z}||_{2} + \text{cos}|| {U}_{h}||_{2}.
\end{equation}
Such a change is effective in reducing the impact of $|| {U}_{h}||_{2}$ on $\eta$, however, it remains that $\gamma_{1}$ may explode.\\




\begin{figure*}[!th]
\begin{center}
\addtolength{\tabcolsep}{-1em}
\scalebox{0.8}{
\begin{tabular}{cc cc cc cc} 

\multicolumn{2}{c}{
\begin{tikzpicture}
\begin{axis}[
    width=5cm, height=5cm,   
    grid = major,
    grid style={dashed, gray!30},
    xmin=1,   
    xmax=1000,  
    ymin=0,   
    ymax=30,  
    axis background/.style={fill=white},
    tick align=outside,
    xtick={1, 250, 500, 750, 1000},
    ytick={1, 5, 10, 15, 20, 25, 30},
    mark repeat={600},
    ylabel near ticks,
    xlabel near ticks,
    xlabel style={text width=5cm},
    xlabel style={align=center},
    xlabel={Epochs},
    domain=0:30,
    restrict y to domain=0:30,
    title=$\eta$
]

\addplot+[mark=none, dash pattern=on 3pt off 4pt on 5pt off 5pt,color=orange,color=blue,line width=1pt] file {resources/dataset/adding_task/ligru_relu/lambda_epoch.txt};

\addplot+[mark=none, color=green,line width=1pt] file {resources/dataset/adding_task/ligru_layernorm/lambda_epoch.txt};

\addplot+[mark=none, color=red,line width=1pt] file {resources/dataset/adding_task/ligru_sin/lambda_epoch.txt};

\addplot+[mark=none, color=brown,line width=1pt] file {resources/dataset/adding_task/ligru_relu_gc_wd/lambda_epoch.txt};

\addplot+[mark=none, dash pattern=on 3pt off 4pt on 5pt off 5pt,color=orange,line width=1pt] file {resources/dataset/adding_task/ligru_sor/lambda_epoch.txt};

\addplot+[only marks, mark size=4pt, color=blue, mark=x, mark options={draw=blue, fill=blue}] coordinates {
		(440,30)
	};
	
\addplot+[only marks, mark size=4pt, color=brown, mark=x, mark options={draw=brown, fill=brown}] coordinates {
		(354,30)
	};

\legend{} 


\end{axis}
\end{tikzpicture}

}

& \multicolumn{2}{c}{
\begin{tikzpicture}
\begin{axis}[
    width=5cm, height=5cm,   
    grid = major,
    grid style={dashed, gray!30},
    xmin = 1,   
    xmax = 1000,  
    ymin = 0,   
    ymax = 1.,  
    axis background/.style={fill=white},
    tick align=outside,
    xtick = {1, 250, 500, 750, 1000},
    ytick = {0, 0.25, 0.5, 0.75, 1, 1.5, 2, 2.5, 3, 3.5, 4, 4.5, 5},
    mark repeat = {600},
    ylabel near ticks,
    xlabel near ticks,
    xlabel style={text width=5cm},
    xlabel style={align=center},
    xlabel={Epochs},
    domain=0:1,
    restrict y to domain=0:1,
    title=MSE
]

\addplot+[mark=none,dash pattern=on 3pt off 4pt on 5pt off 5pt,color=orange, color=blue,line width=1pt,each nth point={25}] file {resources/dataset/adding_task/ligru_relu/loss_epoch.txt};

\addplot+[mark=none, color=green,line width=1pt,each nth point={25}] file {resources/dataset/adding_task/ligru_layernorm/loss_epoch.txt};

\addplot+[mark=none, color=red,line width=1pt,each nth point={25}] file {resources/dataset/adding_task/ligru_sin/loss_epoch.txt};

\addplot+[mark=none, color=brown,line width=1pt,each nth point={25}] file {resources/dataset/adding_task/ligru_relu_gc_wd/loss_epoch.txt};

\addplot+[mark=none, dash pattern=on 3pt off 4pt on 5pt off 5pt,color=orange,line width=1pt,each nth point={25}] file {resources/dataset/adding_task/ligru_sor/loss_epoch.txt};

\addplot+[only marks, mark size=4pt, color=blue, mark=x, mark options={draw=blue, fill=blue}] coordinates {
		(440,1)
	};
	
\addplot+[only marks, mark size=4pt, color=brown, mark=x, mark options={draw=brown, fill=brown}] coordinates {
		(354,0.5)
	};
	
\legend{}

\end{axis}
\end{tikzpicture}

}

& \multicolumn{2}{c}{

\begin{tikzpicture}
\begin{axis}[
    width=5cm, height=5cm,   
    grid = major,
    grid style={dashed, gray!30},
    xmin=1,   
    xmax=1000,  
    ymin=0,   
    ymax=10,  
    axis background/.style={fill=white},
    tick align=outside,
    xtick={1, 250, 500, 750, 1000},
    ytick={0, 2, 4, 6, 8, 10},
    mark repeat={600},
    ylabel near ticks,
    xlabel near ticks,
    xlabel style={text width=5cm},
    xlabel style={align=center},
    xlabel={Epochs},
    domain=0:10,
    restrict y to domain=0:10,
    title={$||U_{h}||_{2}$}
]

\addplot+[mark=none,dash pattern=on 3pt off 4pt on 5pt off 5pt,color=orange, color=blue,line width=1pt] file {resources/dataset/adding_task/ligru_relu/uh_l2_norm_epoch.txt};

\addplot+[mark=none, color=green,line width=1pt] file {resources/dataset/adding_task/ligru_layernorm/uh_l2_norm_epoch.txt};

\addplot+[mark=none, color=red,line width=1pt] file {resources/dataset/adding_task/ligru_sin/uh_l2_norm_epoch.txt};

\addplot+[mark=none, color=brown,line width=1pt] file {resources/dataset/adding_task/ligru_relu_gc_wd/uh_l2_norm_epoch.txt};

\addplot+[mark=none, dash pattern=on 3pt off 4pt on 5pt off 5pt,color=orange,line width=1pt] file {resources/dataset/adding_task/ligru_sor/uh_l2_norm_epoch.txt};

\addplot+[only marks, mark size=4pt, color=blue, mark=x, mark options={draw=blue, fill=blue}] coordinates {
		(442,8.750720977783203)
	};
	
\addplot+[only marks, mark size=4pt, color=brown, mark=x, mark options={draw=brown, fill=brown}] coordinates {
		(356, 3.114022970199585)
	};


\legend{}
\end{axis}
\end{tikzpicture}
} 

&

\multicolumn{2}{c}{

\begin{tikzpicture}

\begin{axis}[
    width=5cm, height=5cm,   
    grid = major,
    grid style={dashed, gray!30},
    xmin=1,   
    xmax=1000,  
    ymin=0,   
    ymax=10,  
    axis background/.style={fill=white},
    tick align=outside,
    xtick={1, 250, 500, 750, 1000},
    ytick={0, 2, 4, 6, 8, 10},
    mark repeat = {600},
    ylabel near ticks,
    xlabel near ticks,
    xlabel style={text width=5cm},
    xlabel style={align=center},
    xlabel={Epochs},
    legend to name=named,
    legend columns=-1,
    restrict y to domain=0:10,
    domain=0:10,
    title={$||U_{z}||_{2}$}
]

\addplot+[mark=none, dash pattern=on 3pt off 4pt on 5pt off 5pt,color=orange,color=blue,line width=1pt] file {resources/dataset/adding_task/ligru_relu/uz_l2_norm_epoch.txt};
\addlegendentry{Standard LiGRU}

\addplot+[mark=none, color=green,line width=1pt] file {resources/dataset/adding_task/ligru_layernorm/uz_l2_norm_epoch.txt};
\addlegendentry{Layer Normalization}

\addplot+[mark=none, color=red,line width=1pt] file {resources/dataset/adding_task/ligru_sin/uz_l2_norm_epoch.txt};
\addlegendentry{Sine Activation}

\addplot+[mark=None,
        color=brown,
        line width=1pt] file {resources/dataset/adding_task/ligru_relu_gc_wd/uz_l2_norm_epoch.txt};
\addlegendentry{GC and WD}

\addplot+[mark=none, dash pattern=on 3pt off 4pt on 5pt off 5pt, color=orange,line width=1pt] file {resources/dataset/adding_task/ligru_sor/uz_l2_norm_epoch.txt};
\addlegendentry{SOR}

\addplot+[only marks, mark size=4pt, color=blue, mark=x, mark options={draw=blue, fill=blue}] coordinates {
		(442, 8.88270378112793)
	};
	
\addplot+[only marks, mark size=4pt, color=brown, mark=x, mark options={draw=brown, fill=brown}] coordinates {
		(356, 5.923626899719238)
	};
	
\end{axis}
\end{tikzpicture}

}


\\

 \multicolumn{8}{c}{ \ref{named} 
} 
\end{tabular}

}


\caption{Results of the adding task with a sequence length of $2,000$ with different Li-GRUs equipped with different methods to prevent gradient exploding to happen (Section \ref{subsec:methods}). $\eta$ is the scale of the gradient that transports the error in time. It must remain close to $1$. The MSE loss function  must be close to $0$. $||U_{z}||_{2}$ and $||U_{h}||_{2}$ shall remain as stable or small as possible. ``GC and WD'' refers to a Li-GRU with weight decay and gradient clipping while ``SOR'' is the soft orthogonal constraint. A small coloured cross indicates that gradients exploded. The layer normalization is the only method that solves the issue by keeping $\eta$ close to $1$, and at the same time, exhibiting the best MSE.}
\vspace{-0.4cm}
\label{fig:toy_task}
\end{center}
\end{figure*}
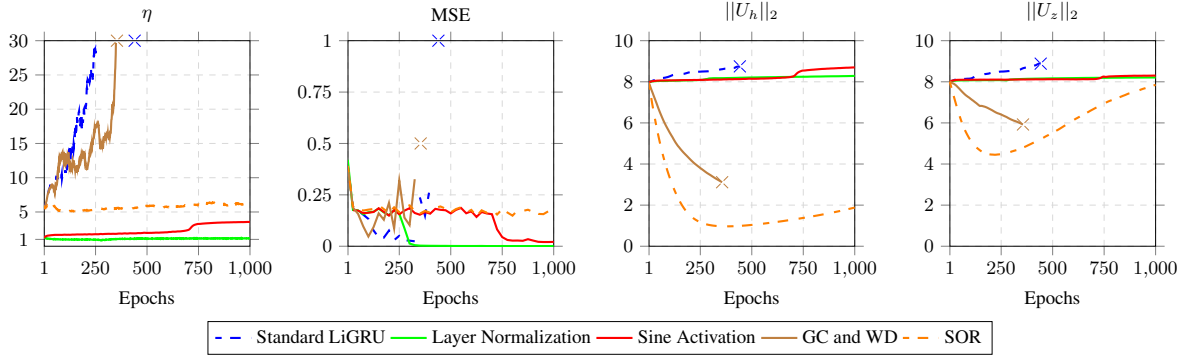

\noindent\textbf{Layer normalization and recurrent weights.} Layer normalization ($\mathcal{LN}$) \cite{ba2016layer} normalizes its input to have zero mean and unit variance. If applied to the product of $U_{*}$ and $h_{t-1}$ in Eq. \ref{eq:relu_ligru} and Eq. \ref{eq:zt_ligru}, it will also reduce the impact of $\gamma_{1}$ on $\eta$, rescale the gradient of $ {U}_{*}$, and accelerate the convergence of the Li-GRU. Introducing the $\mathcal{LN}$ to the original Li-GRU leads to a novel Li-GRU:
\begin{equation}
    z_{t} = \sigma(\mathcal{BN}( {W}_{z}x_{t}) + \mathcal{LN}( {U}_{z}h_{t-1})),
\end{equation}
\begin{equation}
    \tilde{h} = \text{ReLU}(\mathcal{BN}( {W}_{h}x_{t}) + \mathcal{LN}( {U}_{h}h_{t-1})),
\end{equation}
\begin{equation}
    h_{t} = z_{t} \odot h_{t-1} + (1 - z_{t}) \odot \tilde{h}.
\end{equation}
Following the proof from \cite{NEURIPS2019_f8eb278a}, $\eta$ becomes:
\begin{equation}
    \eta = \frac{\gamma_{1}}{4\sigma_{z}}|| {U}_{z}||_{2} + \frac{1}{\sigma_{h}}||  {U}_{h}||_{2}.
\end{equation}
with $\sigma$ the standard deviation of the corresponding gates. In the latter formulation, and motivated by \cite{xu2019understanding}, we decided to get rid of the gain and bias terms of the layer normaliation to even further simplify the architecture. In short, a simple $\mathcal{LN}$ enables $||\partial E_{t}/\partial h_{m}||$ to remain unaffected by the scaling of the recurrent weights but also minimizes the value of $\gamma_{1}$ due to the normalization of the input.






\section{Empirical Evidences}


The empirical validity of the different proposed solutions to the exploding gradients phenomenon is first investigated to properly identify a successor of the Li-GRU named Stabilized Li-GRU (SLi-GRU). Hence, we monitor various metrics including $\eta$ during a task designed to rapidly trigger large gradients (Section \ref{sub:adding_task}). Then, the SLi-GRU is compared to a standard Li-GRU and well-known LSTM networks on three datasets of speech recognition (Section \ref{sub:speech_reco_tasks}). 

\begin{table*}[!ht]

\centering
\caption{Results of different CRDNN equipped with different RNN expressed in terms of word error rate (WER) and character error rate (CER) (\textit{i.e.} lower is better) on the test sets of CommonVoice French, Italian and LibriSpeech. \textit{Inf} refers to infinity \textit{i.e.} exploding gradients. \textbf{Params (M)} is the number of neural parameters. Results in bold are the best results over the experiments. As expected the original Li-GRU exploded on all tasks while the SLi-GRU obtains the best performance even compared to the best LSTM-based CRDNN from SpeechBrain.} \label{table:scaled_datasets} 
\scalebox{0.8}{
\begin{tabular}{cccc|ccc|ccc|ccc}
\hline
\multicolumn{1}{l}{} & \multicolumn{3}{c}{\textbf{CV Italian}} & \multicolumn{3}{c}{\textbf{CV French}} &\multicolumn{3}{c}{\textbf{LS 960 Clean}} &\multicolumn{3}{c}{\textbf{LS 960 Other}}    
\\ \hline
 \textbf{Method} & \textbf{CER \%} & \textbf{WER \%} & 
\textbf{Params (M)} 
  & \textbf{CER \%} & \textbf{WER \%} & 
\textbf{Params (M)} 
 & \textbf{CER \%} & \textbf{WER \%} & \textbf{Params (M)} 
  & \textbf{CER \%} & \textbf{WER \%} & \textbf{Params (M)} 
 \\ \hline
 
\multicolumn{1}{l|}{LSTM}  & 3.50 & 11.06 & \multicolumn{1}{c|}{148.2}  & 5.86 & 14.61 & 148.2 & 1.87 & 3.83  & 173.0 &  6.0 & 10.36  & 173.0\\
\multicolumn{1}{l|}{SLi-GRU} & \textbf{3.05} & \textbf{10.05} &  \multicolumn{1}{c|}{148.1} & \textbf{5.32} & \textbf{13.46} & 148.1 & \textbf{1.27} & \textbf{3.42} & 170.0 & \textbf{3.77} & \textbf{7.82} & 170.0\\
\multicolumn{1}{l|}{LSTM} &  6.82 & 16.28  &  \multicolumn{1}{c|}{50.8} &  7.46 & 18.96 & 50.8 & 2.83 & 4.74 & 94.8 & 8.48 & 12.84 & 94.8\\
\multicolumn{1}{l|}{SLi-GRU} & 3.56 & 11.56 &  \multicolumn{1}{c|}{50.6} & 6.34 & 15.53 & 50.6 & 2.04 & 4.06 & 94.3 & 5.96 & 10.41  & 94.3\\

\hline

\multicolumn{1}{l|}{LiGRU} & \textit{Inf} & \textit{Inf} & \multicolumn{1}{c|}{148.1} & \textit{Inf} & \textit{Inf} & 148.1 & \textit{Inf} & \textit{Inf} & 170.0 & \textit{Inf} & \textit{Inf} & 170.0\\

\multicolumn{1}{l|}{LiGRU} & \textit{Inf} & \textit{Inf} & \multicolumn{1}{c|}{50.6} & \textit{Inf} & \textit{Inf} & 50.6 & \textit{Inf} & \textit{Inf} & 94.3 & \textit{Inf} & \textit{Inf} & 94.3\\

\hline
\end{tabular}
}
\vspace{-0.3cm}
\label{table:speech_reco_tasks}
\end{table*}

\subsection{The adding task}
\label{sub:adding_task}
The adding task \cite{hochreiter1997long} is a synthetic benchmark designed to generate long sequences quickly highlighting exploding gradients. In practice, a neural network must learn to predict the addition of two numbers that are stored in large sequences. Indeed, having long sequences increases the probability of seeing $\partial h_{t}/\partial h_{m}$ explode. Then, the task requires the network to store real values for very long periods, hence giving a convenient way of evaluating the network memory.\\

\noindent\textbf{Task Definition.} Each input sequence is designed as a pair of components. The first component is a real value sampled from a uniform distribution in the range $[0, 1]$ while the second is either $1$ or $0$, and is used as a marker. The goal of the network in this task is to remember the values of the first component of the two elements that have a $1$ in the second component so that the sum can be produced at the output. All input sequences have a length of $T$. The first marker equal to $1.0$ is randomly picked in the range $[0, T/2-1]$ while the other is in the range $[T/2, T-1]$. Using a large value of $T$ (\textit{e.g} $2,000$) is sufficient to create the exploding gradient problem and to challenge the network to do a correct summation. \\


\noindent\textbf{Architecture Details and Network Training.} All models are trained following the same setup so that the effect of the different techniques may be properly compared. The maximum number of epochs is $1,000$, and the number of time steps $T$ is $2,000$. The batch size is $256$. Adam \cite{kingma2014adam} is used as the optimizer with a learning rate of $0.001$. Each Li-GRU configuration used a ReLU if not mentioned in the legend and a single hidden layer with $1,024$ hidden units. The soft orthogonal regularization $\lambda$ is set to $0.001$ after carefully trying multiple values. Weight decay is fixed to $0.001$ as well and the gradient clipping norm threshold is given at $1$. The training loss is the standard mean squared error (MSE) and the recurrent weights follow an orthogonal initialization \cite{saxe2013exact}.\\

\noindent\textbf{Results.} As shown in Fig. \ref{fig:toy_task}, the layer normalization outperformed all the other methods on the principal metrics (i.e $\eta$ and MSE). Not only the $\mathcal{LN}$ keep $\eta$ close to $1$, but also converged faster to an excellent MSE of $0.0005$ whereas the sine activation, \textit{i.e} the second best, achieved an MSE of $0.01$. The sine-based Li-GRU is also slower than its $\mathcal{LN}$ counterpart to converge. Moreover, the value of $\eta$ deviated from $1$ with the sine Li-GRU as the number of epochs reached $700$ indicating the start of an instability in the gradients. The SOR, on the other hand, and despite promising performance to postpone the problem of exploding gradients, was not able to reach a correct MSE due to a negative effect of the constraint applied to the recurrent weights. The standard Li-GRU and its gradient clipping and weight decay variant were not able to prevent the exploding gradient phenomenon and crashed at the same time than the standard Li-GRU. The experiments costed a total of $0.04$ kg of $CO_{2}$ \cite{parcollet:hal-03190119}. \\

\noindent\textbf{Introducing the SLi-GRU.} This analysis demonstrates that the layer normalized Li-GRU, named SLi-GRU thereafter, is the only architecture to: (1) remember past information, hence leading to an excellent MSE score, (2) postpone the gradient exploding problem successfully under tight conditions, and (3) offer a theoretical formulation of $\eta$ that satisfies the need for a high level of stability.



\subsection{Automatic speech recognition experiments}
\label{sub:speech_reco_tasks}

The task of speech recognition is to transcribe the content of an audio signal into a human-readable output. One of the challenges of speech recognition lies in the long input sequences potentially triggering exploding gradients with RNNs. In the following, and to illustrate the latter issue, we compare our newly introduced SLi-GRU to Li-GRU and LSTM neural networks on three different datasets including CommonVoice French, Italian and LibriSpeech.\\ 

\noindent\textbf{Datasets.} 
Three datasets with different complexities and sizes are considered to evaluate the models: LibriSpeech \cite{panayotov2015librispeech}, CommonVoice French (version 8.0) \cite{ardila2020common}, and CommonVoice Italian (version 8.0) \cite{ardila2020common}. These datasets are primer choices as they exhibit long sequences making the training challenging for the SLi-GRU while remaining extremely competitive as the community has been extensively investigating them. The utterances of CommonVoice (CV) are obtained from volunteers all around the world. The French set (CV-fr) contains $460$K utterances ($671$ hours) with different accents, and more than $16$K participants. The train set consists of $620.59$ hours, while both validation and test sets contain $25.52$, and $25.52$ hours of speech respectively. The Italian set (CV-it), is relatively small compared to CV-fr. It contains $208$, $24$, $26$ hours training, validation, and test data. Finally, LibriSpeech (LS) is a corpus of $960$ hours of read and clean English speech. \\ 

\noindent\textbf{Architecture Details and Network Training.}
We trained an LSTM \cite{greff2016lstm}, a Li-GRU, and a SLi-GRU with two different scenarios on a single and popular end-to-end (E2E) ASR architecture with all datasets: an encoder-decoder CTC-Attention based CRDNN coming from the SpeechBrain toolkit \cite{ravanelli2021speechbrain}. The encoder is composed of three distinct parts: a VGG-like features extractor, a bidirectional RNN, and a deep dense neural network. This is combined with a location-aware attentive GRU decoder jointly trained with the CTC loss. The two different scenarios that are considered for the Li-GRU are: (1) a \textit{low-budget model (LB)}, as claimed by the original Li-GRU authors, we wished to verify if a smaller Li-GRU could achieve state-of-the-art performance. In this particular case, the ASR system has $50.6$M and $94.3$M parameters on CV and LS respectively; (2) a \textit{high-budget model (HB)}, with the Li-GRU matching approximately the same number of neural parameters than the best LSTM-based CRDNN from SpeechBrain (\textit{i.e} $148$M and $170$M parameters). Hyperparameters and neural architectures vary across the different datasets and are extensively described in the corresponding \textit{SpeechBrain} recipes \cite{ravanelli2021speechbrain} (commit hash \textit{ee50231}). A recurrent language model (LM) trained with the LS language modelling resources is coupled via shallow fusion for LibriSpeech. No LM is used with CV. \\

\noindent\textbf{Results.} First, it is worth underlining that the standard Li-GRU did not even succeed in lasting more than one epoch on all tasks as the gradient exploding problem formally demonstrated in this work simply hits hard empirically as well. Such a problem is solved with the introduction of our SLi-GRU that was able to proceed for the entire training hence proving that the recurrent layer normalization is an effective solution to approach exploding gradients in RNNs. Then, with a fixed size of $148$M and $170$M parameters, the SLi-GRU outperformed the previous state-of-the-art CRDNN model from SpeechBrain on the three datasets with WER of $7.82\%$, $3.42\%$, $13.46\%$ and $10.05\%$ compared to $10.36\%$, $3.83\%$, $14.61\%$ and $11.06\%$ for the LSTM on the LS other and clean, and CV French and Italian datasets respectively. Even more interestingly, the \textit{LB} SLi-GRU solely equipped with a third and half of the neural parameters on CV and LS reached promising performance as the WER only degraded by $0.425$ in average compared to the \textit{HB} LSTM over the three datasets. Furthermore, the \textit{LB} SLi-GRU reduced the WER by $21\%$ in average against the \textit{LB} LSTM. Following the original findings of the Li-GRU \cite{ravanelli2018light}, it seems clear that our SLi-GRU represents an interesting alternative to LSTM for speech recognition in \textit{HB} and \textit{LB} scenarios. Finally, we estimate that at least $39.3$ kg of $CO_{2}$ have been emitted to produce these results \cite{parcollet:hal-03190119}. 





\subsection{Efficiency perspectives}

The original Li-GRU made available to the community did not leverage any optimization scheme at training time. Hence, it introduced an extremely large overhead in iteration time, even compared to much more complex LSTM networks that directly benefit from a CUDA implementation. As we wish our method to impact widely and positively the community, we decided to implement a custom CUDA version of the SLi-GRU so that anyone may use it with large datasets without suffering from excessive training times. The latter contribution is five times faster and change the time-complexity from a quadratic to a linear regime, compared to the default PyTorch implementation. This contribution replaces the previous Li-GRU implementation in the well-known SpeechBrain toolkit.\\
\section{Conclusion}
We introduced the SLi-GRU, a theoretically grounded and empirically validated stabilized version of the original Li-GRU. Following an analytical demonstration, we showed that applying a layer-wise normalization to the recurrent gate of the SLi-GRU is sufficient to prevent the exploding gradients phenomenon. During the conducted experiments, the SLi-GRU outperformed the previous state-of-the-art relying on LSTM from the SpeechBrain toolkit in three automatic speech recognition tasks. Furthermore, we also released an optimized version of the SLi-GRU further reducing the required training time by a factor of five. 


\bibliographystyle{IEEEbib}
\bibliography{refs, strings}

\begin{thebibliography}{10}

\bibitem{karita2019comparative}
Shigeki Karita, Nanxin Chen, Tomoki Hayashi, Takaaki Hori, Hirofumi Inaguma,
  Ziyan Jiang, Masao Someki, Nelson Enrique~Yalta Soplin, Ryuichi Yamamoto,
  Xiaofei Wang, et~al.,
\newblock ``A comparative study on transformer vs rnn in speech applications,''
\newblock in {\em 2019 IEEE Automatic Speech Recognition and Understanding
  Workshop (ASRU)}. IEEE, 2019, pp. 449--456.

\bibitem{watanabe2017hybrid}
Shinji Watanabe, Takaaki Hori, Suyoun Kim, John~R Hershey, and Tomoki Hayashi,
\newblock ``Hybrid ctc/attention architecture for end-to-end speech
  recognition,''
\newblock {\em IEEE Journal of Selected Topics in Signal Processing}, vol. 11,
  no. 8, pp. 1240--1253, 2017.

\bibitem{graves2006connectionist}
Alex Graves, Santiago Fern{\'a}ndez, Faustino Gomez, and J{\"u}rgen
  Schmidhuber,
\newblock ``Connectionist temporal classification: labelling unsegmented
  sequence data with recurrent neural networks,''
\newblock in {\em Proceedings of the 23rd international conference on Machine
  learning}, 2006, pp. 369--376.

\bibitem{graves2014towards}
Alex Graves and Navdeep Jaitly,
\newblock ``Towards end-to-end speech recognition with recurrent neural
  networks,''
\newblock in {\em International conference on machine learning}. PMLR, 2014,
  pp. 1764--1772.

\bibitem{ravanelli2021speechbrain}
Mirco Ravanelli, Titouan Parcollet, Peter Plantinga, Aku Rouhe, Samuele
  Cornell, Loren Lugosch, Cem Subakan, Nauman Dawalatabad, Abdelwahab Heba,
  Jianyuan Zhong, et~al.,
\newblock ``Speechbrain: A general-purpose speech toolkit,''
\newblock {\em arXiv preprint arXiv:2106.04624}, 2021.

\bibitem{watanabe2018espnet}
Shinji Watanabe, Takaaki Hori, Shigeki Karita, Tomoki Hayashi, Jiro Nishitoba,
  Yuya Unno, Nelson-Enrique~Yalta Soplin, Jahn Heymann, Matthew Wiesner, Nanxin
  Chen, et~al.,
\newblock ``Espnet: End-to-end speech processing toolkit,''
\newblock {\em Proc. Interspeech 2018}, pp. 2207--2211, 2018.

\bibitem{amodei2016deep}
Dario Amodei, Sundaram Ananthanarayanan, Rishita Anubhai, Jingliang Bai, Eric
  Battenberg, Carl Case, Jared Casper, Bryan Catanzaro, Qiang Cheng, Guoliang
  Chen, et~al.,
\newblock ``Deep speech 2: End-to-end speech recognition in english and
  mandarin,''
\newblock in {\em International conference on machine learning}. PMLR, 2016,
  pp. 173--182.

\bibitem{ravanelli2018light}
Mirco Ravanelli, Philemon Brakel, Maurizio Omologo, and Yoshua Bengio,
\newblock ``Light gated recurrent units for speech recognition,''
\newblock {\em IEEE Transactions on Emerging Topics in Computational
  Intelligence}, vol. 2, no. 2, pp. 92--102, 2018.

\bibitem{loshchilov2017decoupled}
Ilya Loshchilov and Frank Hutter,
\newblock ``Decoupled weight decay regularization,''
\newblock in {\em ICLR}, 2019.

\bibitem{pascanu2013difficulty}
Razvan Pascanu, Tomas Mikolov, and Yoshua Bengio,
\newblock ``On the difficulty of training recurrent neural networks,''
\newblock in {\em International conference on machine learning}. PMLR, 2013,
  pp. 1310--1318.

\bibitem{vorontsov2017orthogonality}
Eugene Vorontsov, Chiheb Trabelsi, Samuel Kadoury, and Chris Pal,
\newblock ``On orthogonality and learning recurrent networks with long term
  dependencies,''
\newblock in {\em International Conference on Machine Learning}. PMLR, 2017,
  pp. 3570--3578.

\bibitem{werbos1990backpropagation}
Paul~J Werbos,
\newblock ``Backpropagation through time: what it does and how to do it,''
\newblock {\em Proceedings of the IEEE}, vol. 78, no. 10, pp. 1550--1560, 1990.

\bibitem{arjovsky2016unitary}
Martin Arjovsky, Amar Shah, and Yoshua Bengio,
\newblock ``Unitary evolution recurrent neural networks,''
\newblock in {\em International conference on machine learning}. PMLR, 2016,
  pp. 1120--1128.

\bibitem{NIPS2017_f2fc9902}
Sekitoshi Kanai, Yasuhiro Fujiwara, and Sotetsu Iwamura,
\newblock ``Preventing gradient explosions in gated recurrent units,''
\newblock in {\em Advances in Neural Information Processing Systems}, I.~Guyon,
  U.~Von Luxburg, S.~Bengio, H.~Wallach, R.~Fergus, S.~Vishwanathan, and
  R.~Garnett, Eds. 2017, vol.~30, Curran Associates, Inc.

\bibitem{ioffe2015batch}
Sergey Ioffe and Christian Szegedy,
\newblock ``Batch normalization: Accelerating deep network training by reducing
  internal covariate shift,''
\newblock in {\em International conference on machine learning}. PMLR, 2015,
  pp. 448--456.

\bibitem{cho2014properties}
Kyunghyun Cho, Bart van Merri{\"e}nboer, Dzmitry Bahdanau, and Yoshua Bengio,
\newblock ``On the properties of neural machine translation: Encoder--decoder
  approaches,''
\newblock {\em Syntax, Semantics and Structure in Statistical Translation}, p.
  103, 2014.

\bibitem{NEURIPS2019_f8eb278a}
Lu~Hou, Jinhua Zhu, James Kwok, Fei Gao, Tao Qin, and Tie-Yan Liu,
\newblock ``Normalization helps training of quantized lstm,''
\newblock in {\em Advances in Neural Information Processing Systems},
  H.~Wallach, H.~Larochelle, A.~Beygelzimer, F.~d\textquotesingle
  Alch\'{e}-Buc, E.~Fox, and R.~Garnett, Eds. 2019, vol.~32, Curran Associates,
  Inc.

\bibitem{mikolov2012statistical}
Tom{\'a}{\v{s}} Mikolov et~al.,
\newblock ``Statistical language models based on neural networks,''
\newblock {\em Presentation at Google, Mountain View, 2nd April}, vol. 80, no.
  26, 2012.

\bibitem{parascandolo2016taming}
Giambattista Parascandolo, Heikki Huttunen, and Tuomas Virtanen,
\newblock ``Taming the waves: sine as activation function in deep neural
  networks,'' 2017.

\bibitem{ba2016layer}
Jimmy~Lei Ba, Jamie~Ryan Kiros, and Geoffrey~E Hinton,
\newblock ``Layer normalization,''
\newblock {\em stat}, vol. 1050, pp. 21, 2016.

\bibitem{xu2019understanding}
Jingjing Xu, Xu~Sun, Zhiyuan Zhang, Guangxiang Zhao, and Junyang Lin,
\newblock ``Understanding and improving layer normalization,''
\newblock {\em Advances in Neural Information Processing Systems}, vol. 32,
  2019.

\bibitem{hochreiter1997long}
Sepp Hochreiter and J{\"u}rgen Schmidhuber,
\newblock ``Long short-term memory,''
\newblock {\em Neural computation}, vol. 9, no. 8, pp. 1735--1780, 1997.

\bibitem{kingma2014adam}
Diederik~P Kingma and Jimmy Ba,
\newblock ``Adam: A method for stochastic optimization,''
\newblock in {\em ICLR (Poster)}, 2015.

\bibitem{saxe2013exact}
Andrew~M. Saxe, James~L. McClelland, and Surya Ganguli,
\newblock ``Exact solutions to the nonlinear dynamics of learning in deep
  linear neural networks,''
\newblock {\em CoRR}, vol. abs/1312.6120, 2014.

\bibitem{parcollet:hal-03190119}
Titouan Parcollet and Mirco Ravanelli,
\newblock ``{The Energy and Carbon Footprint of Training End-to-End Speech
  Recognizers},''
\newblock in {\em Proc. Interspeech 2021}, 2021, pp. 4583--4587.

\bibitem{panayotov2015librispeech}
Vassil Panayotov, Guoguo Chen, Daniel Povey, and Sanjeev Khudanpur,
\newblock ``Librispeech: an asr corpus based on public domain audio books,''
\newblock in {\em 2015 IEEE international conference on acoustics, speech and
  signal processing (ICASSP)}. IEEE, 2015, pp. 5206--5210.

\bibitem{ardila2020common}
Rosana Ardila, Megan Branson, Kelly Davis, Michael Kohler, Josh Meyer, Michael
  Henretty, Reuben Morais, Lindsay Saunders, Francis Tyers, and Gregor Weber,
\newblock ``Common voice: A massively-multilingual speech corpus,''
\newblock in {\em Proceedings of the 12th Language Resources and Evaluation
  Conference}, 2020, pp. 4218--4222.

\bibitem{greff2016lstm}
Klaus Greff, Rupesh~K Srivastava, Jan Koutn{\'\i}k, Bas~R Steunebrink, and
  J{\"u}rgen Schmidhuber,
\newblock ``Lstm: A search space odyssey,''
\newblock {\em IEEE transactions on neural networks and learning systems}, vol.
  28, no. 10, pp. 2222--2232, 2016.

\end{thebibliography}

\end{document}